# Skyrmion crystals in frustrated Shastry-Sutherland magnets

J. H. Yu, W. H. Li, Z. P. Huang, J. J. Liang, J. Chen, D. Y. Chen, Z. P. Hou, and M. H. Qin[*]

*Guangdong Provincial Key Laboratory of Quantum Engineering and Quantum Materials, Institute for Advanced Materials, South China Academy of Advanced Optoelectronics, South China Normal University, Guangzhou 510006, China*

**[Abstract]** The phase diagrams of the frustrated classical spin model with Dzyaloshinskii-Moriya (DM) interaction on the Shastry-Sutherland (S-S) lattice are studied by means of Monte Carlo simulation. For ferromagnetic next-nearest-neighboring ($J_2$) interactions, the introduced exchange frustration enhances the effect of the DM interaction, which enlarges the magnetic field-range with the skyrmion lattice phase and increases the skyrmion density. For antiferromagnetic $J_2$ interactions, the so-called $2q$ phase (two-sublattice skyrmion crystal) and the spin-flop phase are observed in the simulated phase diagram, and their stabilizations are closely dependent on the DM interaction and $J_2$ interaction, respectively. The simulated results are qualitatively explained from the energy landscape, which provides useful information for understanding the intriguing phases in S-S magnets.

Keywords: Skyrmion, frustration, Shastry-Sutherland model

PACS numbers: 12.39.Dc, 75.25.+z, 75.30.Kz

[*]Electronic mail: qinmh@scnu.edu.cn

# I. Introduction

A skyrmion is a topological defect with a vortex-like spin structure where the spins point in all directions wrapping a sphere.[1,2] During the past a few years, skyrmions have drawn extensive attention for their potential applications in future memory technology, especially considering their outstanding merits such as the nanoscale sizes,[3,4] the topological protection, and the ultralow critical drive-currents[5]. Specifically, skyrmions have been discovered experimentally in chiral magnets including MnSi[1,6], MnGe[7] and FeGe[8], and have been proven theoretically to be stabilized by the competition among the ferromagnetic (FM) exchange, Dzyaloshinskii-Moriya (DM)[9,10], and Zeeman couplings in the presence of thermal fluctuations. Subsequently, several effective methods for manipulating skyrmions using spin-polarized current[11,12], magnetic field gradients[13-15], microwave field[16], and temperature gradient[17-20] have been revealed experimentally and/or theoretically.

Furthermore, strong interfacial DM interactions can be induced at heavy metal (HM)/FM interfaces in films such as Pt/Co/MgO[21] heterostructure and Pt/Co/Ta[22] thin film due to broken inversion symmetry and strong spin-orbit coupling, contributing to the stabilization of Néel-type magnetic skyrmions even at room temperature ($T$). More recently, in the IrMn/CoFeB/MgO heterostructure, it has been reported that the exchange bias at the antiferromagnetic (AFM)/FM interface can stabilize the skyrmions even in the absence of external magnetic field ($B$).[23] More interestingly, the DM interaction at the interface can be significantly enhanced by a factor of 7 through increasing the IrMn thickness, which is very meaningful for spintronic applications.

In addition, several earlier works on finding new magnetic systems with skyrmions have been reported. For example, several frustrated magnets such as the triangular antiferromagnets have been predicted to host skyrmion lattice phase. Instead of DM interaction, the competition between the nearest-neighbor (NN) and next-nearest-neighbor (NNN) interactions is essential for the stabilization of the skyrmions, as discussed in detail in the earlier works.[24-30] Most recently, the first observation of skyrmionic magnetic bubbles formed at room temperature has been reported experimentally in the frustrated kagome $Fe_3Sn_2$ magnet,[31] well confirming the earlier predictions.

It is well known that finding materials with skyrmions stabilized in a wide $B$ range is essential both in basic physical research and for potential applications. In some extent, frustrated magnets such as Shastry-Sutherland (S-S) magnets may be promising candidates. On the one hand, there are real materials such as rare-earth tetraborides $R$B$_4$ ($R$ = Tb, Dy, Ho, Tm, etc.) are famous representatives of S-S magnets. In these materials, DM interactions could be induced by spin-orbit coupling between rare-earth ions and/or by structural inversion asymmetry through S-S magnet film design, and may play an important role in stabilizing non-collinear phases. For example, in the S-S Kondo lattice model, the DM interactions have been suggested to be necessary for the emergence of the chiral spin structures.[32] Thus, skyrmion lattice phase could be expected to be stabilized in S-S magnets with DM interactions. On the other hand, the exchange frustration has been proven to be important in understanding the intriguing magnetization behaviors observed in these materials.[33,34] More importantly, it is strongly expected that the DM effect could be enhanced by the effective reduction of the neighboring spin interactions due to the exchange frustration, and in turn further stabilizes skyrmions. As a matter of fact, similar phenomenon has been observed in the frustrated spin model on the square lattice, and the enlargement of the $B$ range with the skyrmions by considering the interaction frustration has been reported.[35] Thus, there is still an urgent need to elucidate complex magnetic orders in frustrated S-S magnets with DM interactions, although the multi-step magnetization behaviors in these materials have been extensively investigated in recent years.[36-38]

In this work, we study phase diagrams of a frustrated spin model with DM interaction on the Shastry-Sutherland lattice. For FM NNN interaction ($J_2$), the introduced exchange frustration enhances the effect of the DM interaction, leading to an enlargement of the magnetic field-range with the skyrmion lattice phase and an increase of the skyrmion density. For AFM $J_2$ interaction, the 2$q$ phase (two-sublattice skyrmion crystal) and the spin-flop phase are suggested to be stabilized by the DM interaction and $J_2$ interaction, respectively. The simulated results are qualitatively explained from the energy landscape.

## II. Model and method

In the presence of additional DM interaction, the Hamiltonian is given by

$$H = -\sum_{i,\delta_n} J_n \mathbf{S}_i \cdot \mathbf{S}_j - \sum_{\langle i,j \rangle} \mathbf{D}_{ij} \cdot \left( \mathbf{S}_i \times \mathbf{S}_j \right) - \sum_i \mathbf{B} \cdot S_i^z \quad (1)$$

The first term is the exchange interactions where $j = i + \delta_n$ and $\delta_n$ connects site $i$ and its $n$-th nearest neighbor sites with $n = 1, 2, 3$. Here, $J_n$ are the exchange couplings and $\mathbf{S}_i$ represents the Heisenberg spin with unit length on site $i$. The second term describes the interfacial DM interaction arises from structural inversion asymmetry along thin-film normal direction with the DM vectors $D_{ij}$ are shown in Fig. 1. The last term is the Zeeman coupling with $B$ applied along the $z$ direction.

Our simulation is performed on an $N = 24 \times 24$ S-S lattice with period boundary conditions using the standard Metropolis algorithm and the parallel tempering algorithm.[39] We take an exchange sampling after every 10 standard Monte Carlo (MC) steps. Generally, the initial $5 \times 10^5$ MC steps are discarded for equilibrium consideration and another $5 \times 10^5$ MC steps are retained for statistical averaging of the simulation. To explore the phases in the system, besides the well-known magnetization $M$ and magnetic susceptibility $\chi$, we also characterize the spin structures by performing the Fourier transform

$$\mathbf{A}_\mathbf{k} = \frac{1}{N} \sum_i \mathbf{S}_i \exp(-i\mathbf{k} \cdot \mathbf{r}_i) \quad (2)$$

where $\mathbf{r}_i$ is the spatial coordinate of site $i$, and then calculating the intensity profile $|\mathbf{A}_\mathbf{k}|^2$. In addition, we also calculate the topological winding number $Q_{sk}$ defined by

$$Q_{sk} = \frac{1}{4\pi} \iint \mathbf{S}_i \cdot \left( \partial_x \mathbf{S}_i \times \partial_y \mathbf{S}_i \right) dxdy \quad (3)$$

to characterize the number of skyrmions in the system.[40]

### III. Simulation results and discussion

#### A. Enhanced skyrmion lattice phase

First, the effect of exchange frustration on the stabilization of skyrmions at low temperature $T = 0.01$ is investigated. Without loss of generality, we fix FM $J_2 = 0.5$ and $D = 0.73$, and modulate the frustration by introducing AFM couplings $J_1 = 2J_3 = -\alpha$. Figs. 2(a) and 2(b) present respectively the magnetization $M(B)$ curves and susceptibility $\chi(B)$ curves for various $\alpha$. In the absence of exchange frustration at $\alpha = 0$, two discontinuities are clearly observed in the simulated magnetization $M(B)$ curve, indicating two subsequent phase transitions with the increase of $B$. It is noted that the ground state under zero $B$ for $\alpha = 0$ is the spiral phase with wave vector $\mathbf{k} = \arctan(D/\sqrt{2} J_2)$ (1, 1). The phase transition from the spiral phase to the skyrmion lattice phase occurs when $B$ increases to a critical value. Subsequently, the FM order is stabilized under strong enough $B$.

The two critical fields can be reasonably estimated from the positions of the two peaks in the simulated susceptibility curve. With the increase of $\alpha$, the second peak significantly shifts toward the high $B$ side, while the first peak is almost invariant, clearly demonstrating an enlargement of the $B$-range with the skyrmion lattice phase. This phenomenon can be understood from two aspects. On the one hand, when AFM $J_1$ and $J_3$ couplings are considered, the energy loss from the $J_1$ coupling energy $H_1$ and the $J_3$ coupling energy $H_3$ due to the phase transition from the skyrmion lattice phase to the FM phase is introduced, resulting in the increase of the saturation field. Thus, rather than the FM phase, the skyrmion lattice phase can be further stabilized when $\alpha$ is increased, as clearly shown in our simulations. On the other hand, $H_1$ and $H_3$ are almost invariant in the phase transition from the spiral phase to the skyrmion lattice phase (the corresponding results are not shown here), and the first critical field is rather stable as $\alpha$ increases. As a result, the $B$-range with the skyrmion lattice phase is obviously enlarged, which is also confirmed in the simulated $Q_{sk}(B)$ curves for various $\alpha$ given in Fig. 2(c).

Fig. 2(d) presents the simulated phase diagram in the ($\alpha$, $B$) parameter plane as well as the corresponding winding numbers. For a fixed $B$, $Q_{sk}$ increases with the increasing $\alpha$, indicating an increase/decrease of the density/size of the skyrmion, as clearly shown in Figs. 3(a) and 3(b) which give the spin configuration and the Bragg intensity $|A_\mathbf{k}|^2$ under $B = 0.55$ for $\alpha = 0$ and $\alpha = 0.4$, respectively. It is noted that for $\alpha = 0$, the size of skyrmion is mainly determined by the $J_2/D$ value, i.e., a larger $J_2/D$ results in a smaller size under a fixed $B$. The consideration of the competing $J_1$ and $J_3$ interactions effectively reduces the neighboring spin interactions and enhances the importance of the DM interaction, resulting in the increase/decrease of the density/size of the skyrmion. Thus, our work clearly demonstrates that both the *B*-range and the size of the skyrmion can be effectively modulated by tuning the magnitude of the frustration in S-S magnets. Moreover, the chirality of the skyrmions is invariant due to the DM interaction, which has been confirmed in our simulations.

**B. Stabilization of the 2*q* phase and the spin-flop phase**

Most recently, the 2*q* phase has been observed in the square-lattice antiferromagents, which shows strong similarity to skyrmions in FM films. In this part, we investigate spin orders for AFM $J_2$ under applied magnetic field, and pay particular attention to the effect of $J_2$ on phase diagram at low *T*. The simulated phase diagram in the ($J_2$, *B*) parameter plane at *T* = 0.01 for FM $J_1 = 2J_3 = 0.5$ is presented in Fig. 4, in which the boundaries are similarly estimated from the positions of the peaks in $\chi(B)$ curves. Four phases: the AFM-spiral phase, the 2*q* phase, the spin-flop phase, and the FM phase exist in the phase diagram, and detailed spin configurations and results analysis are given below.

It is noted that spins on the square/S-S lattice tend to form a two sublattice structure. Under low *B*, the AFM-spiral phase is stabilized, in which the staggered magnetization forms the spiral order, as clearly shown in Fig. 5(a) which presents a snapshot (top) and spin configurations on the sublattices A and B (bottom) of the AFM-spiral phase for $J_2 = 0.4$ under *B* = 0.4. The two interpenetrating spirals on sub-lattice A and B are with a same ordering wave vector, as clearly demonstrated in the intensity of the spin structure factor (insert in Fig. 5(a)). When *B* is increased to a critical value, the phase transition from the AFM-spiral phase to the 2*q* phase is occurred. In the 2*q* phase of which typical configuration under *B* = 1.8 is given in Fig. 5(b), the sublattice spin configurations show similar characteristics as FM spins in skyrmion lattice phase, except that the skyrmion density maps are checker-board-like rather than hexagonal. Moreover, the critical field increases as the magnitude of $J_2$ increases. With the further increase of *B*, the 2*q* phase is directly replaced by the FM phase for $J_2 < 0.15$ to save the Zeeman energy $H_{zee}$. For a large $J_2$ ($J_2 > 0.15$), the spin-flop phase sandwiched between the 2*q* phase and the FM phase is observed. The same as earlier report, all the spins in the spin-flop phase have a same nonzero *z* component and their *xy* components form the AFM structure, as clearly shown in Fig. 5(c) which gives the spin configuration of the spin-flop phase under *B* = 2.2. In addition, the third transition $B$ (black triangular points in the phase diagram) significantly increases with the increase of $J_2$, showing an enlargement of the $B$-range with the spin-flop phase.

The simulated phase diagram can be well understood from the energy landscape. Fig. 6(a) gives the calculated $B$-dependence of $H_1$, the $J_2$ coupling energy $H_2$, $H_3$, the DM interaction energy $H_{DM}$, and $H_{zee}$ for $J_2 = 0.25$. The corresponding magnetization and susceptibility curves are also presented in Fig. 6(b) to help one to understand the results easier. Under small $B$ ($B <$ 1.2), the AFM-spiral phase is stabilized by the cooperation of the exchange and DM interactions, similar to that in square-lattice antiferromagnets. Within certain $B$ range (1.2 < $B$ < 1.5), the energy loss from $H_2$ and $H_{DM}$ due to the transition from the AFM-spiral phase to the $2q$ phase is overtaken by the energy gain mainly from $H_{zee}$, resulting in the stabilization of the $2q$ phase. With the further increase of $B$, the sizes of the sublattice skyrmions are decreased to increase the magnetization and to save $H_{zee}$, although the corresponding results are not shown here. Moreover, when $B$ increases to ~ 1.5, the transition from the $2q$ phase to the spin-flop phase is occurred to save $H_1$, $H_3$, and $H_{zee}$ in the expense of $H_2$ and $H_{DM}$. The $z$ component of the spins of the spin-flop phase linearly increases with $B$ until the emergence of the FM phase, as clearly shown in the simulated magnetization curve in Fig. 6(b).

At last, we tend to discuss the transitions of the critical fields with the enhancement of the $J_2$ interaction. It is noted that $H_2$ increases in the successive two or three phase transitions, as confirmed in Fig. 6(c) which gives the simulated local energies for $J_2 = 0.4$. Thus, the critical fields increase when $J_2$ is increased, as shown in Fig. 6(d) which gives the magnetization and susceptibility as function of $B$ for $J_2 = 0.4$. Furthermore, for a fixed $J_2$, the energy loss from $H_2$ due to the phase transition from the initial spin-flop phase to the FM phase is rather larger than the energy loss due to the phase transition from the initial $2q$ phase to the spin-flop phase. As a result, the third critical $B$ increases more quickly than the second critical one with the increasing $J_2$, resulting in an enlargement of the $B$-range with the spin-flop phase, as shown in the simulated phase diagram.

Importantly, it is clearly shown that $H_{DM}$ significantly increases with $B$ in the $2q$ phase, while is invariant in the spin-flop phase. Therefore, the $2q$ phase can be enhanced by the DM interaction, resulting in an enlargement of the $B$-range with the $2q$ phase when $D$ is increased. This phenomenon has been also confirmed in our simulations, and the corresponding results are shown in Fig. 7 which gives the simulated phase diagram in the ($D$, $B$) parameter plane for $J_2 = -0.4$. It is clearly shown that the third transition $B$ is invariant and the spin-flop phase is gradually replaced by the $2q$ phase with the increase of $D$.

## IV. Conclusion

In summary, we have studied the phase diagrams of the frustrated spin model with the DM interaction on the Shastry-Sutherland lattice using Monte Carlo simulation. For FM next-nearest-neighboring ($J_2$) interaction, the frustration induced by considering the competing exchange interactions enhances the effect of the DM interaction, resulting in not only an enlargement of the field-range with the skyrmion lattice phase but also an increase of the skyrmion density. For AFM $J_2$ interaction, the so-called $2q$ phase and the spin-flop phase are observed in the simulated phase diagram, and the magnetic field range with the $2q$/spin-flop phase is enlarged with the increase of the $J_2$/DM interaction. The simulated results are discussed in detail from the energy landscape.

**Acknowledgements**:

The work is supported by the National Key Projects for Basic Research of China (Grant No. 2015CB921202), and the Natural Science Foundation of China (Grant No. 51332007), and the Science and Technology Planning Project of Guangdong Province (Grant No. 2015B090927006), and the Natural Science Foundation of Guangdong Province (Grant No. 2016A030308019).

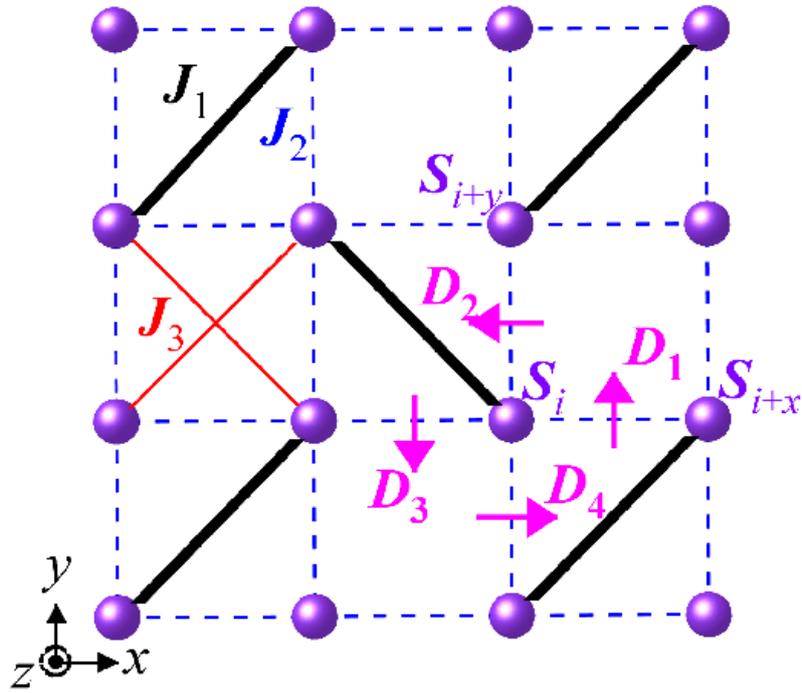

Figure. 1. (color online) Effective model on the Shastry-Sutherland lattice model with the exchange and additional DM interactions. The DM vectors are depicted by the pink arrows.

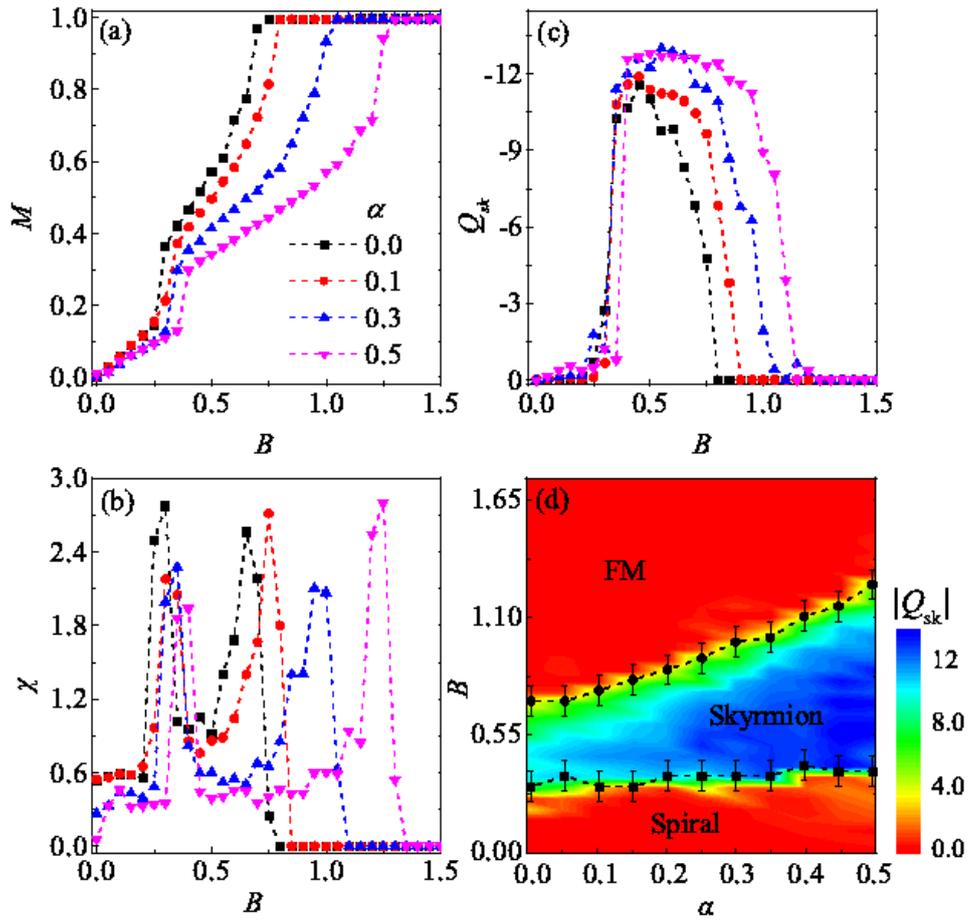

Figure. 2. (color online) The calculated (a) magnetization $M$, (b) magnetic susceptibility $\chi$, and (c) the topological winding number $Q_{sk}$ as functions of $B$ for various $\alpha$ at $T$ = 0.01 and $J_2$ = 0.5. (d) The summarized phase diagram and the value of $Q_{sk}$ in the ($\alpha$, $B$) parameter plane.

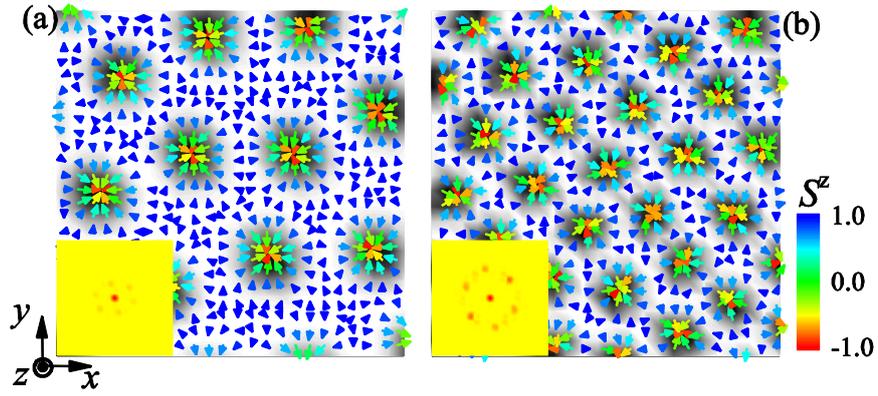

Figure. 3. (color online) Typical snapshot of the spin configurations of the skyrmion lattice phase for $J_2 = 0.5$ at $T = 0.01$ for (a) $(\alpha, B) = (0.0, 0.55)$, and (b) $(0.4, 0.55)$. The inserts show the Bragg intensity $|\mathbf{A_k}|^2$.

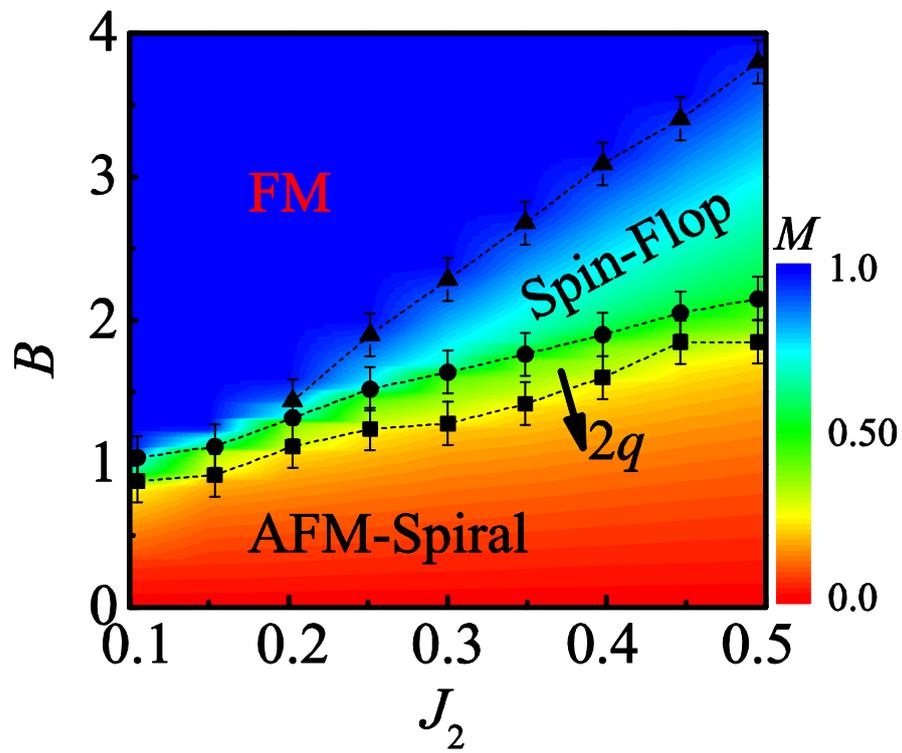

Figure. 4. (color online) The simulated phase diagram in the ($-J_2$, $B$) plane at $T = 0.01$ for FM $J_1 = 2J_3 = 0.5$.

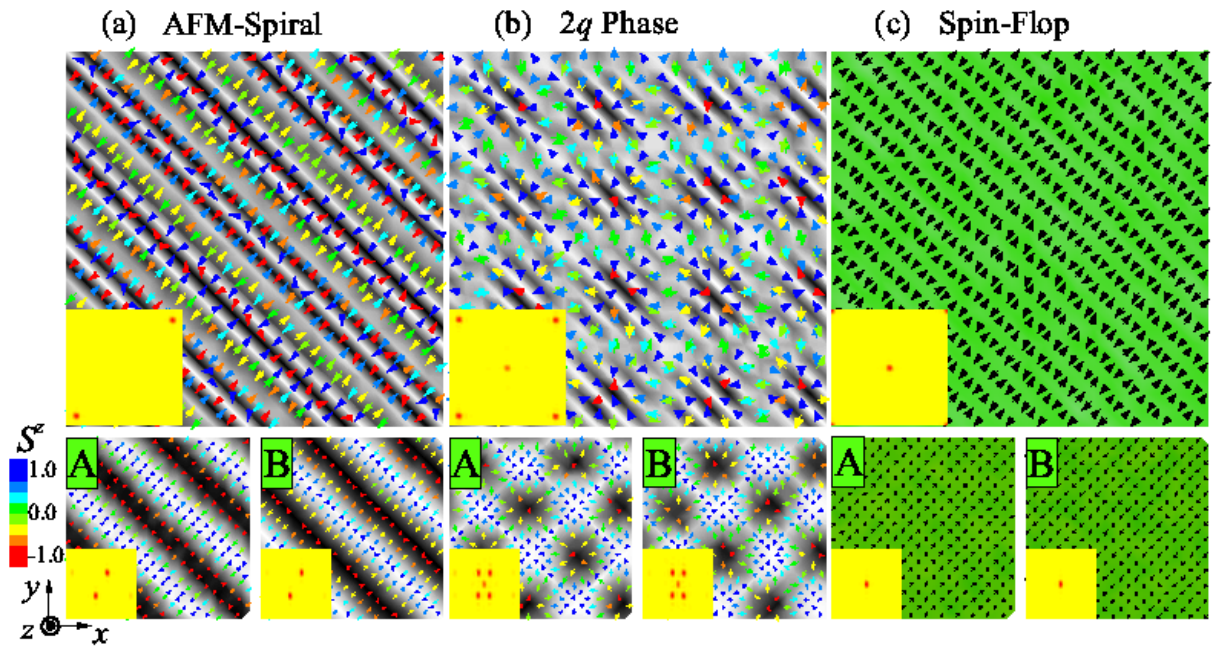

Figure. 5. (color online) Typical snapshot of the spin configuration of (a) the AFM-spiral phase under $B = 0.4$, and (b) the $2q$ phase under $B = 1.8$, and (c) the spin-flop phase under $B = 2.2$ for $J_2 = 0.4$. The two sub-lattice spin structures (bottom) and the Bragg intensities $|A_k|^2$ (inserts) are also presented.

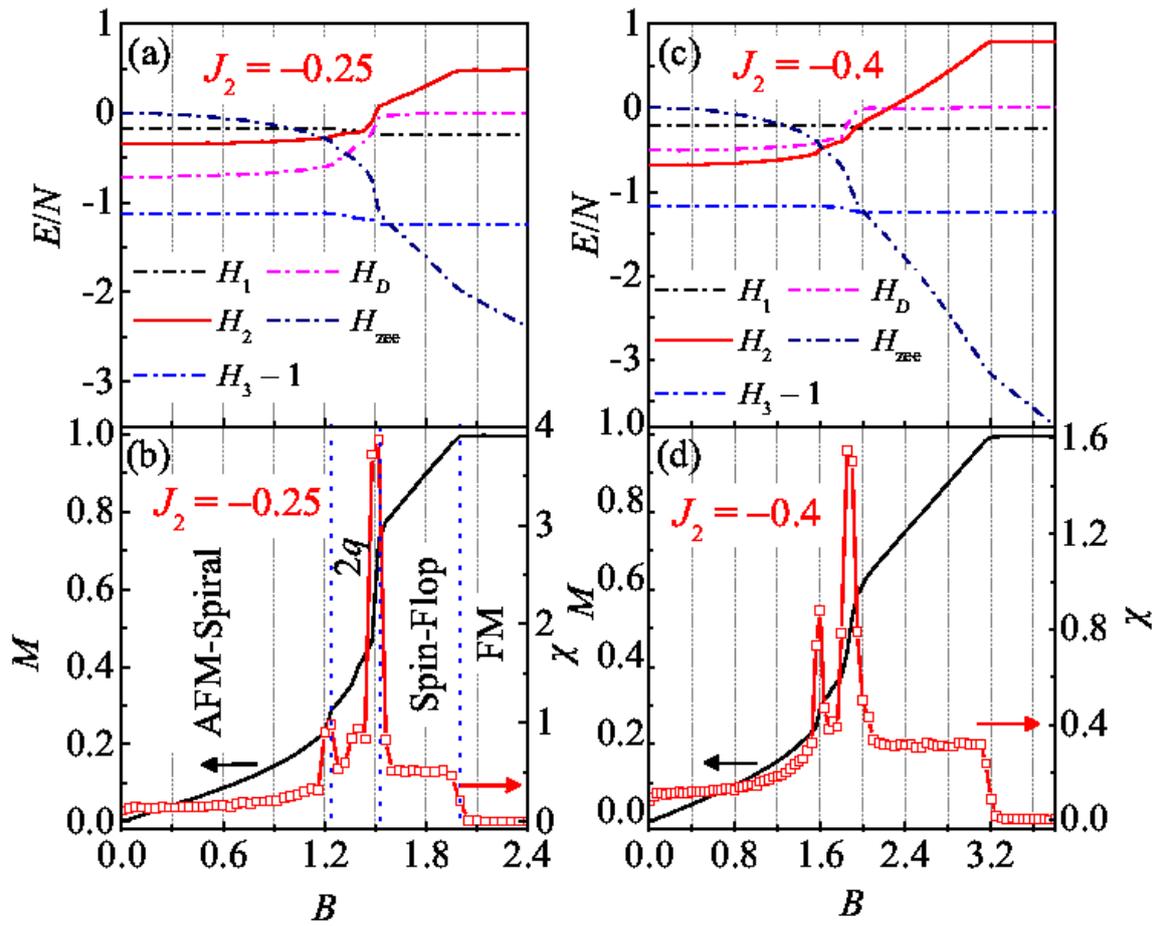

Figure. 6. (color online) The local energies as functions of $B$ at $T = 0.01$ for (a) $J_2 = 0.25$, and (c) $J_2 = 0.4$. (b) and (d) are the correspondingly magnetization and magnetic susceptibility curves.

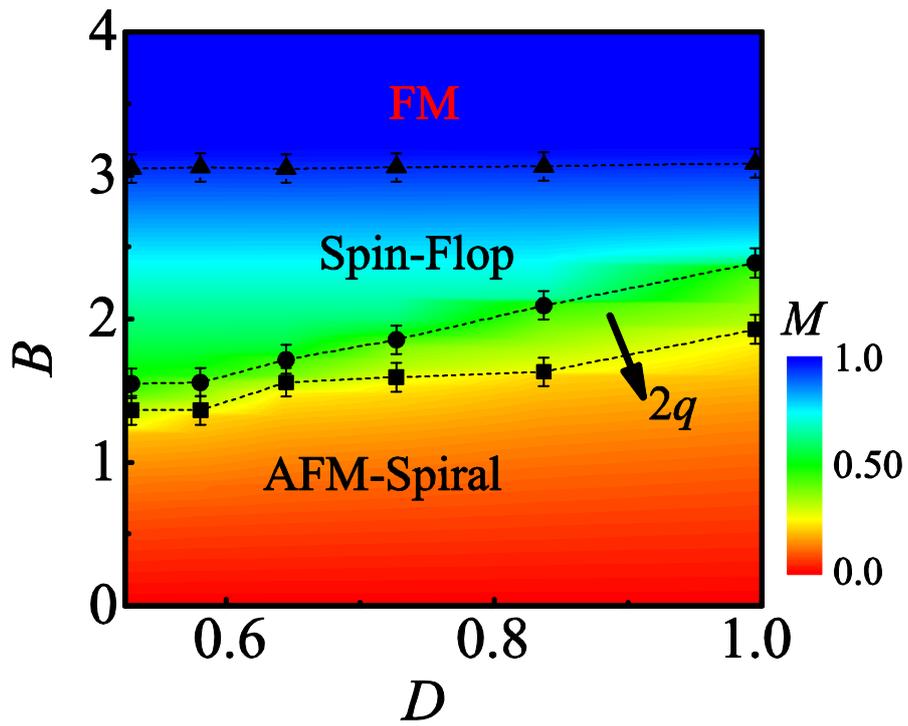

Figure. 7. (color online) The simulated phase diagram in the $(D, B)$ plane at $T = 0.01$ for FM $J_1$ = $2J_3$ = 0.5 and $J_2$ = −0.4.